# Giant Enhancement of Nonlinear Harmonic Generation in a Silicon Topological Photonic Crystal Nanocavity Chain

*Qingchen Yuan, Linpeng Gu, Liang Fang, Xuetao Gan*, Zhigang Chen, Jianlin Zhao**

Q. Yuan, L. Gu, L. Fang, Prof. X. Gan, Prof. J. Zhao
Key Laboratory of Light Field Manipulation and Information Acquisition, Ministry of Industry and Information Technology, and Shaanxi Key Laboratory of Optical Information Technology, School of Physical Science and Technology, Northwestern Polytechnical University, Xi'an 710129, China
E-mail: *xuetaogan@nwpu.edu.cn; jlzhao@nwpu.edu.cn*

Prof. Z. Chen
The MOE Key Laboratory of Weak-Light Nonlinear Photonics, TEDA Applied Physics Institute and School of Physics, Nankai University, Tianjin 300457, China

Strongly enhanced third-harmonic generation (THG) by the topological localization of an edge mode in a Su-Schrieffer-Heeger (SSH) chain of silicon photonic crystal nanocavities is demonstrated. The edge mode of the nanocavity chain not only naturally inherits resonant properties of the single nanocavity, but also exhibits the topological feature with mode robustness extending well beyond individual nanocavity. By engineering the SSH nanocavities with alternating strong and weak coupling strengths on a silicon slab, we observe the edge mode formation that entails a THG signal with three orders of magnitude enhancement compared with that in a trivial SSH structure. Our results indicate that the photonic crystal nanocavity chain could provide a promising on-chip platform for topology-driven nonlinear photonics.

## 1. Introduction

Taking the inspiration from the quantum Hall effects and topological insulators discovered in condensed matter physics, topological photonics has turned into a rapidly emerging field of research.[1-5] Its main hallmark is the emergence of topologically protected edge states at the interface between photonic structures with distinct topological invariants.[1,2] Such edge states present a solid immunity to local distortions during optical transport,[3-10] which promotes the development of topological platforms for implementing stable and reliable photonic devices such as single-mode low threshold topological lasers[11-19] and chirality-selective routers.[4,7,8,20-24] When optical nonlinearities are taken into account, novel phenomena and advanced functionalities arise, including nonlinearity-induced topological transitions, topological bandgap solitons and robust single-photon sources.[25-34] For instance, relying on spontaneous four-wave-mixing and other nonlinear processes, topologically protected quantum states including the correlated biphoton pairs and entangled states have been proposed and demonstrated.[29-31] Based on the Kerr effect, self-trapped soliton edge modes are predicted under local topological phase transition with high pump power, promising for tunable filters and isolators with eliminated backscattering.[32] In addition, nonlinear harmonic generations by the edge modes of topological photonic structures have also been reported, which enable advanced nonlinear optical imaging of nanostructures with superior contrast and sensitivity.[33,34] Therefore, nonlinear topological photonics is expected to form a favorable ground with novel functionalities for photonic applications.

One of the prototypical and most popular topological photonic systems is the so-called one-dimensional Su-Schrieffer-Heeger (SSH) model formed by a chain of coupling elements under alternating strong and weak coupling strengths, which can support topological edge modes.[35] In photonics, the first SSH lattice and associated edge modes were realized in an optically induced dimer chain of coupled waveguides, where trivial and non-trivial superlattices were readily reconfigured by different terminations of the quasi-periodic photonic structures.[36] The stationary edge modes of the non-trivial SSH lattices have been widely employed to achieve topological lasing by constructing dimer chains of microrings, nanodisks, and photonic crystal nanocavities.[16-19,37] Interestingly, the introduction of nonlinear effects into the photonic SSH lattices has led to a host of novel phenomena, including for example actively controlled topological zero modes, nonlinear spectral tuning, and nonlinear control of PT symmetry and non-Hermitian topological states.[26,38-43] In particular, a pioneering work about the nonlinear harmonic generation in photonic SSH structure was reported by Kruk *et. al*, which is composed of a zigzag array of silicon nanodisks with Mie resonance.[34] The nanoscale silicon nanodisks, as the coupling element of the SSH structure, promise the compact footprint and flexible design of the structure by moving the nanodisks individually. Topological localization of the electric field at the edge of the zigzag array provides multifold enhancement of the third harmonic generation (THG), while enjoys the topological robustness of the edge modes against perturbations.

In this work, we demonstrate strongly enhanced THG in silicon by exciting a topological edge mode of an SSH lattice formed by coupled photonic crystal nanocavities (PCNCs). We show that, in such an SSH structure, the edge mode naturally inherits resonant properties of a single PCNC, while it also manifests the topological feature with robustness extending well beyond individual nanocavity. Compared with silicon nanodisks,[34] PCNCs fabricated in a silicon slab have resonance modes with much higher quality ($Q$) factors, which are promising for narrower resonance linewidth and stronger light-matter interaction. In addition, it is more reliable to construct on-chip optoelectronic devices with PCNCs, including modulators, photodetectors, etc.[44-46] These attributes enable the SSH structure of PCNCs to have potentials in chip-integrated topological photonic circuits. From the fabricated structures, we obtain a strong silicon THG signal from the non-trivial topological edge mode, which is enhanced by more than three orders of magnitude compared to that from a trivial SSH structure.

## 2. Device Design

**Figure 1**a displays the schematic of the enhanced THG driven by the topological edge mode of a silicon SSH structure. PCNCs as the coupling elements are arranged to form two sublattices ($A_j$ and $B_j$) with hopping coupling strengths of $\gamma_1$ and $\gamma_2$. The finite-sized SSH structure has nine coupled PCNCs, which are separated by one or three rows of air-holes to guarantee their alternating coupling strengths ($\gamma_1 < \gamma_2$). The employed PCNCs here are formed by removing three adjacent air-holes and shifting the nearest neighbor two air-holes outwards

in a hexagonal photonic crystal lattice, i.e. the L3 PCNCs.[47]

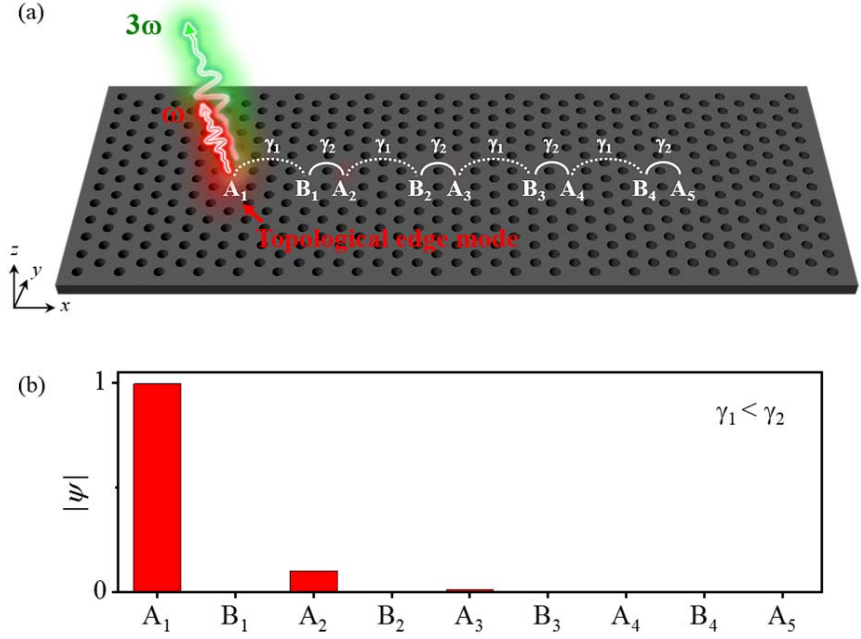

**Figure 1.** a) Schematic of the finite-sized SSH structure formed by nine coupled PCNCs ($A_1B_1A_2B_2…A_5$) with alternating weak and strong coupling strengths ($\gamma_1<\gamma_2$) to support enhanced THG by the topological edge mode. b) Wave function ($|\psi|$) of the topological edge mode based on the tight-binding model.

In the SSH model with odd numbers of sites, the Hamiltonian could be expressed as

$$H_{odd} = \sum_{j=1}^{(N-1)/2} \left[\left(\gamma_1 A_j^\dagger + \gamma_2 A_{j+1}^\dagger\right) B_j\right] + h.c. \qquad (1)$$

Here, $N$ is the number of sites, which is chosen as 9 in this work. By diagonalizing the Hamiltonian, when $\gamma_1<\gamma_2$, one can find an edge mode and its wave function ($|\psi|$) dominants at the left terminus, as shown in Figure 1b. The edge mode here could be predicted by the topological invariant in an infinite topological system, called Zak phase ($\theta_{Zak}$) in the one-dimensional topological system, that is

$$\theta_{Zak} = i\int_{-\pi}^{\pi} dk \langle u_k | i\partial_k | u_k \rangle = \begin{cases} \pi, \gamma_{intra} < \gamma_{inter} \\ 0, \gamma_{intra} > \gamma_{inter} \end{cases} \qquad (2)$$

where $u_k$ is the Bloch wave function. $\gamma_{intra}$ and $\gamma_{inter}$ are intra- and inter-coupling strengths of the dimer unit $A_jB_j$, which respectively correspond to $\gamma_1$ and $\gamma_2$ in Figure 1a.[48] For an SSH chain formed by an infinite number of dimer units $A_jB_j$, the intra-coupling strength $\gamma_{intra}$ is weaker than the inter-coupling strength $\gamma_{inter}$, and the system is consequently in a topological non-trivial phase with $\theta_{Zak}=\pi$. In contrast, if one considers the dimer $B_jA_{j+1}$ as a constitutional unit, the intra-coupling strength of the dimer is stronger than the inter-coupling strength, and consequently the chain turns into a system with a topological trivial phase of $\theta_{Zak}=0$. Therefore, at the interface between the non-trivial chain and surrounding trivial system, such as the terminal site $A_1$ in Figure 1a, there would happen a topological phase transition ($\Delta\theta_{Zak}=\pi$), which supports a topological edge mode.[48]

**3. Results and Discussions**

The proposed SSH structure is fabricated on a silicon-on-insulator substrate with a 220 nm thick top silicon layer. The hexagonal lattice of the photonic crystal has a period of $a=465$ nm and an air-hole radius of $r=0.29a$. This design supports the resonance modes around near-infrared spectral range with high $Q$ factors. After the electron beam lithography and inductively coupled plasma dry etching, the formed structures in the silicon slab are air-suspended by undercutting the buried oxide layer, which ensures the effective vertical confinements of the resonance modes. **Figure 2**a and 2b display scanning electron microscopy (SEM) images of the fabricated single PCNC and non-trivial SSH structure formed by nine coupled PCNCs. To carry out the control experiment, a trivial SSH structure is fabricated as well, which has ten coupled PCNCs, as shown in Figure 2c.

The fabricated devices are characterized by measuring their vertical light scattering using a cross-polarization microscope.[49] A supercontinuum laser is employed as the broadband excitation light source, which is focused by an objective lens (50X, NA=0.42) to a spot size of ~2 μm over the cavity region. This ensures efficient in-situ excitations and detections of the edge modes.[33,34] The vertical scatterings of the resonance modes from the cavities are then collected by the same objective lens and finally analyzed by a spectrometer mounted with an InGaAs camera. To facilitate the study of position-dependence of the edge mode, the devices are mounted on a two-dimensional piezo-actuated stage to realize its spatial movement with respect to the laser focusing point.

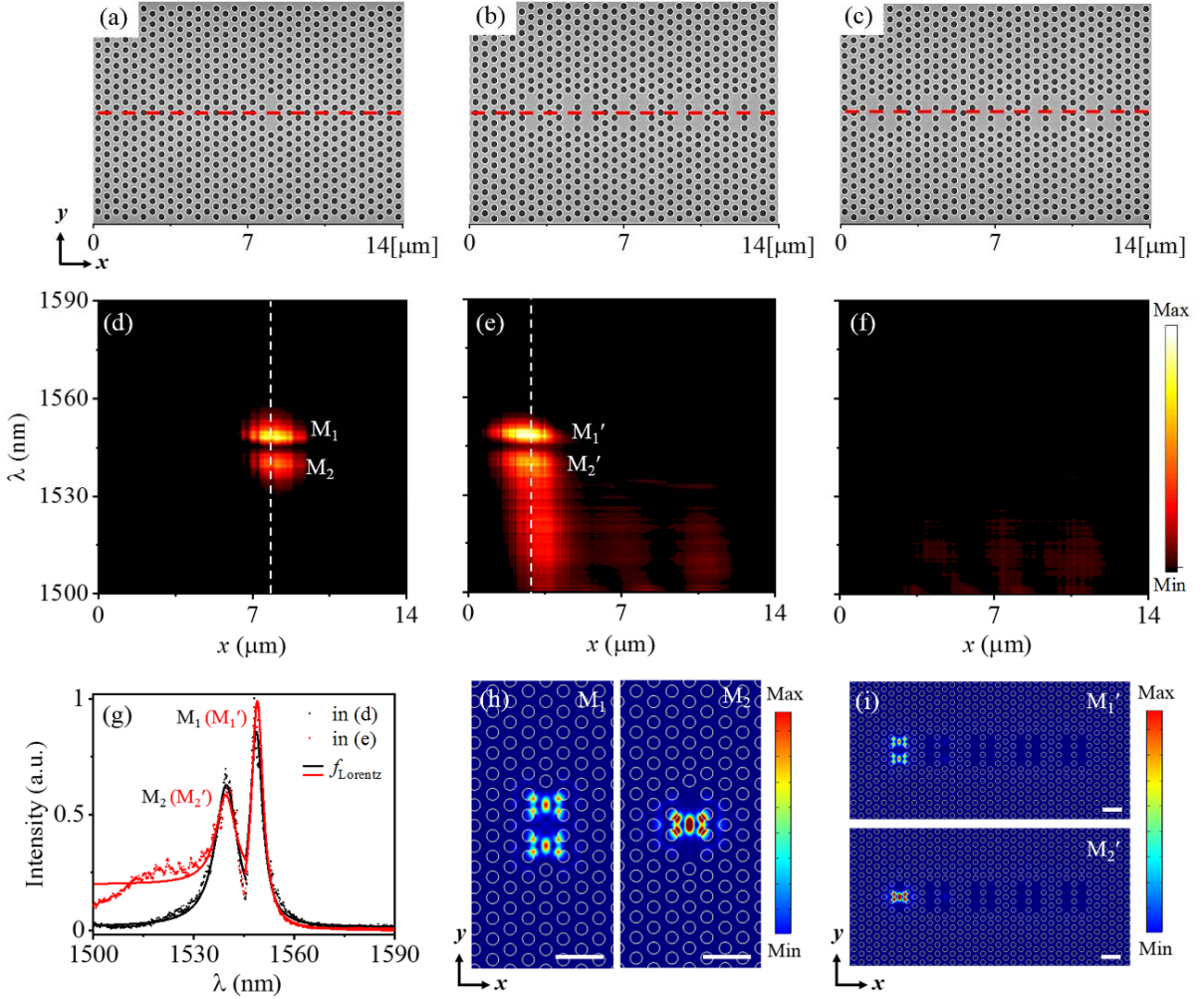

**Figure 2.** a-c) SEM images of the a) single PCNC, b) non-trivial SSH structure and c) trivial SSH structure. d-f) Measured position-dependent scattering spectra along their centerlines in the *x*-axis of the three structures shown in a-c), respectively. g) Scattering spectra of the single PCNC and non-trivial SSH structure marked by the white dashed lines in d) and e), where two scattering peaks are successively marked as $M_1$ ($M_1'$) and $M_2$ ($M_2'$) for the single PCNC (non-trivial SSH structure). The solid lines are their Lorentzian fitting curves. h,i) Simulated electric

field distributions ($|E|^2$) of the two resonance modes in the h) single PCNC and i) non-trivial SSH structure. Scale bar: 1 μm.

Figure 2d displays the measured scattering spectra from the single PCNC as it is spatially moved along the centerline in the *x*-axis, as indicated by the red dashed line in Figure 2a. Since the employed cross-polarization microscope only collects the scattering light with polarization perpendicular to the polarization of the incident laser, there should be no collected scattering signal over the photonic crystal lattice or unpatterned silicon slab. When the incident laser focuses on the cavity region, the light coupled into the cavity excites the resonance mode, whose far-field scattering has multiple polarization components. The scattering light from the resonance mode could therefore be collected and detected by the cross-polarization microscope. As a consequence, the position-dependent scattering spectra shown in Figure 2d represent resonance peaks when the incident laser is focused on the cavity region. These two resonance peaks are then verified from the mode simulations, as discussed below.

The position-dependent scattering spectra from the non-trivial SSH structure are acquired as well by moving it along the *x*-axis, as shown in Figure 2e. Though there are nine PCNCs, scattering peaks are only observed from the left terminal nanocavity $A_1$, matching well with the theoretical result in Figure 1b. The two scattering peaks have similar central wavelengths as those measured from the single PCNC shown in Figure 2d, i.e., the edge mode of SSH structure inherits the resonant property of its constitutive element. In comparison, the position-dependent scattering spectra of the trivial SSH structure shown in Figure 2c are displayed in Figure 2f. Different from the result obtained from the non-trivial SSH structure (see Figure 2e), there is no detectable scattering peak.

To study the resonance modes of the single PCNC and the non-trivial SSH structure in detail, in Figure 2g, we plot the scattering spectra from the interesting locations of the structures, which are indicated by the white dashed lines in Figure 2d and 2e. To assist the discussion, the two scattering peaks are successively marked as $M_1$ ($M_1'$) and $M_2$ ($M_2'$) from long to short wavelength for the single PCNC (the non-trivial SSH structure). Using Lorentzian fittings, in the single PCNC, the $M_1$ mode locates at 1548.72 nm with a $Q$ factor of 340, and the $M_2$ mode locates at 1539.75 nm with a $Q$ factor of 190. From the non-trivial SSH structure, the resonance wavelength of $M_1'$ mode undergoes a red-shift of 0.33 nm with respect to the $M_1$ mode, and the $Q$ factor is improved by 30. The resonance wavelength of $M_2'$ mode has no shift compared to the $M_2$ mode, and the $Q$ factor is improved by 60.

The above experiment results are further verified by the mode simulations based on the finite element technique. Though there are only two resonance modes experimentally observed from the single PCNC, six resonance modes are obtained numerically, which is consistent with the previously reported results.[50] According to the spacing between the resonance wavelengths as well as the polarization attributes of their far-field scatterings, the experimentally observed two peaks are recognized to have the electric field distributions ($|E|^2$) shown in Figure 2h.[50,51] The other four resonance modes are not observed in the experiment, which could be attributed to their low coupling efficiencies in the cross-polarization microscope with the configuration of far-field coupling.[52,53] From the simulation, the resonance wavelengths of $M_1$ and $M_2$ modes are 1538.91 nm and 1529.08 nm, and their $Q$ factors are 580 and 600, respectively. The derivations of the resonance wavelengths and $Q$ factors are arisen from the fabrication errors. Resonance modes of the non-trivial SSH structure with nine coupled PCNCs are then solved as well. Corresponding to the six resonance modes of the single PCNC, six topological edge modes are obtained from the SSH structure. For the experimentally observed two edge modes, which are formed by the couplings of $M_1$ mode and $M_2$ mode respectively, their electric field distributions ($|E|^2$) are displayed in Figure 2i. Strongly localized modes are observed only at the left terminal nanocavity $A_1$. The field profiles at $A_1$ are the same as their counterparts of the single PCNC.

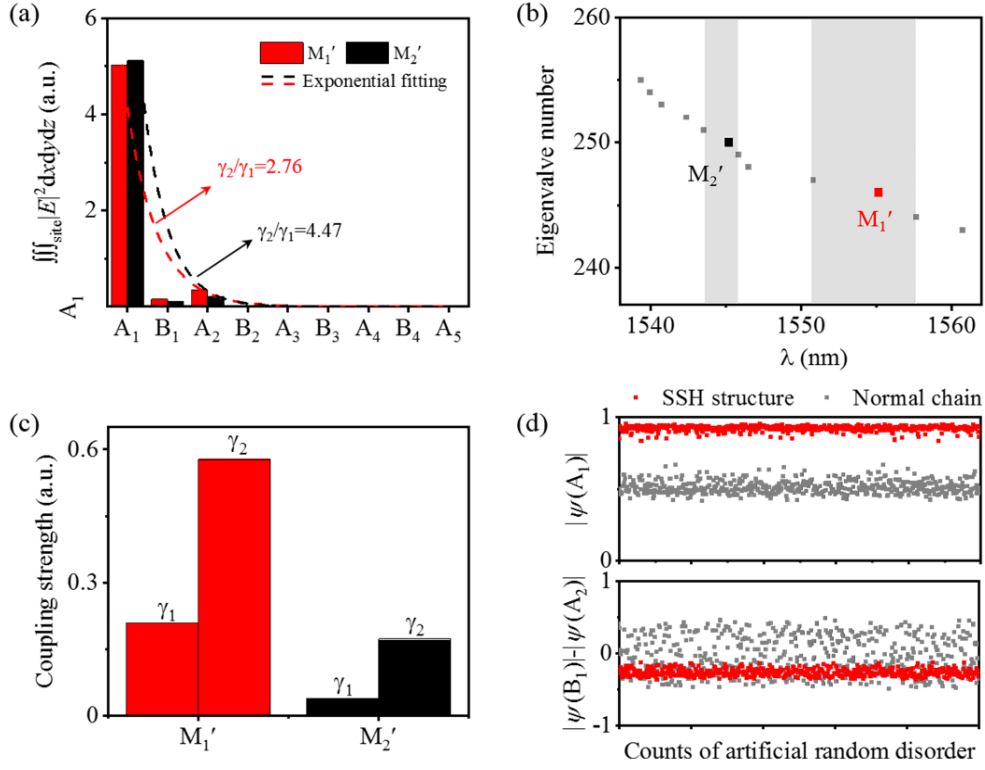

**Figure 3.** a) Electromagnetic energy ($\iiint_{site}|E|^2 dxdydz$) stored at each nanocavity site for modes $M_1'$ and $M_2'$, where the ratio $\gamma_2/\gamma_1$ is fitted from the exponential curve $(\gamma_2/\gamma_1)^{-N}$ along the hopping nanocavities $A_j$. b) Simulated eigenvalues of the non-trivial SSH structure, where the shadow areas are the mode gaps related to the coupling strengths. c) Calculated $\gamma_1$ and $\gamma_2$ from their ratios and mode gaps. d) Localized wave functions at the terminal nanocavity $A_1$ ($|\psi(A_1)|$) (top panel) and differential localized wave functions between the nearest neighbor nanocavity $B_1$ and the next-nearest neighbor nanocavity $A_2$ ($|\psi(B_1)|-|\psi(A_2)|$) (bottom panel) as functions of random disorder for the SSH structure (red dots) and a normal chain with only one single coupling strength $\gamma=(\gamma_1+\gamma_2)/2$ (grey dots), indicating the robustness of the topological edge mode.

Apart from the almost same field profiles in the cavity regions of the single PCNC and the SSH structure, there exists exponentially decaying tail along sites in the SSH structure, which is closely related to the coupling strengths of the specific resonance modes. We integrate electromagnetic energy stored at each nanocavity site using a series of self-defined cuboids ($\iiint_{site}|E|^2 dxdydz$), which covers as much of the electric field profile as the software allows, as shown in **Figure 3**a. Different from the wave function distribution in Figure 1b with no energy at any site $B_j$, there is certain electromagnetic energy redistributed into these "dark" sites, such as $B_1$, which could be attributed to the inevitable perturbation from simulation mesh. Considering that the energies stored at the "dark" sites are weak, the electromagnetic energy envelope analysis of the edge mode still follows the exponential analytical solution from the standard SSH model.[16] The electromagnetic energy envelope of hopping "bright" sites $A_j$ decays as $(\gamma_2/\gamma_1)^{-N}$, where $N$ is the site number counted from the edge. As shown in Figure 3a, using the exponential fitting, $\gamma_2/\gamma_1$ of modes $M_1'$ and $M_2'$ are recognized as 2.76 and 4.47, respectively. Compared to those of mode $M_1'$, $\gamma_2/\gamma_1$ of mode $M_2'$ is higher because the field profile under almost the same sum of electromagnetic field is more compact.

During the eigenvalue calculations of the resonance modes in the simulation, edge modes appear among the mode gaps of bulk modes, as indicated by the grey areas in Figure 3b. Note the edge modes are generally off the true center of the mode gap on account of the resonance wavelengths deviation of the nine PCNCs. In the infinite SSH model, the mode gap is $2|\gamma_1-\gamma_2|$.[48] In the finite SSH model with nine sites, by diagonalizing the Hamiltonian in triple symmetric diagonal 9×9 matrix form in Equation (1), the mode gap is specifically equal to $2\sqrt{\gamma_1^2 + \gamma_2^2 - \frac{\sqrt{5}+1}{2}\gamma_1\gamma_2}$, which is slightly larger because of the looser confinement from the finite number sites to edge mode. Combining with their ratios estimated from Figure 3a, $\gamma_1$ and $\gamma_2$ of the two resonance modes are finally determined in Figure 3c. It is demonstrated that there are larger coupling strengths of $M_1'$ for its more widespread field distribution, where there is more profile overlapping with the adjacent nanocavity. The larger coupling strength could provide more designs with different patterns to form an SSH model with the maintained robustness.

Based on the calculated resonance wavelengths and coupling strengths, we artificially introduce random disorders in the resonance wavelengths below 1% and coupling strengths below 10% in tight-binding model to investigate the robustness of the obtained topological edge modes.[19] For comparison, a normal chain with only one single coupling strength $\gamma=(\gamma_1+\gamma_2)/2$ is also considered. The wave functions localized at the terminal nanocavity $A_1$ ($|\psi(A_1)|$) of the edge mode in the SSH structure and the normal chain are recorded with varied disorder strengths, as shown in the top panel of Figure 3d. The $|\psi(A_1)|$ of the edge mode in the normal chain (grey dots) fluctuates considerably. Differently, that of the edge mode $M_1'$ in the SSH structure (red dots) is more steady. Meanwhile, the differential wave function localized between the nearest neighbor nanocavity $B_1$ and the next-nearest neighbor nanocavity $A_2$ ($|\psi(B_1)|-|\psi(A_2)|$) are also calculated, as shown in the bottom panel of Figure 3d. The sign of $|\psi(B_1)|-|\psi(A_2)|$ is the key to determine whether an edge mode is a topological edge mode of the SSH model.[48] When the sign is negative, the obtained edge mode maintains the quasi-"dark" sublattice distribution and belongs to a topological edge mode. Otherwise, it manifests a general edge mode without topological character. As the bottom panel of Figure 3d shown, the sign of $|\psi(B_1)|-|\psi(A_2)|$ in the normal chain (grey dots) changes between positive and negative depending on the disorder, while that in the SSH structure (red dots) always keeps negative. Therefore, the edge mode in the non-trivial SSH structure is immune against the random structural disorders under topological protection. However, when the SSH chain degrades into a normal chain, the supported edge will lose the topological character and become impressionable. With the mean value of the localized wave functions in Figure 3d, the edge mode $M_1'$ remains 98% robustness profiting from the topological protection. The same calculation is implemented for the edge mode $M_2'$ as well, showing a mean value of 93%.

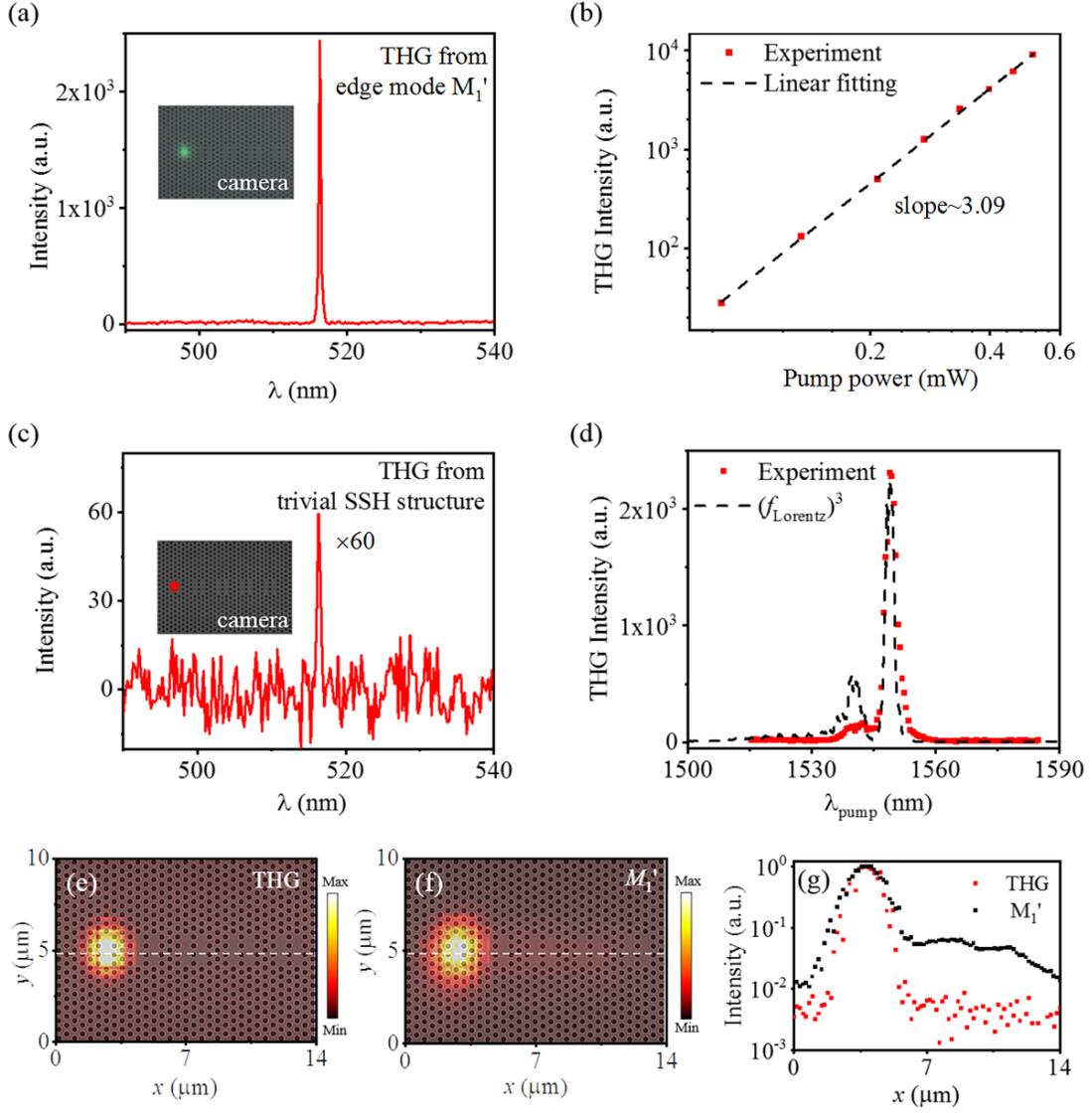

**Figure 4.** a) Measured THG spectrum from the non-trivial SSH structure under the pump laser at 1549.05 nm (resonance wavelength of the $M_1'$ mode). Inset shows the acquired THG image using a visible CMOS camera. b) Log-log plot of the power-dependence of the THG signal, showing a line fitting with a slope about 3.09. c) THG spectrum obtained from the trivial SSH structure under the same pump condition used in a). d) Pump wavelength dependence of the THG signal from the non-trivial SSH structure, where the black dashed line is the third power of the scattering spectrum shown in Figure 2g. e,f) Spatial profiles of the e) scattering THG signal and f) scattering pump laser (edge mode $M_1'$), which are superimposed over the SEM image of the SSH structure. g) Position-dependence of THG signal and edge mode $M_1'$ along the centerline in the $x$-axis.

PCNC has been well considered as one of the optical resonators with the highest factor of $Q/V$, where $V$ is the mode volume. It therefore could be employed to realize strong light-matter interaction due to the strongly localized electric field of the resonance mode. A variety of nonlinear optical processes have been realized based on a single PCNC, such as second harmonic generation, THG, optical parametric oscillator, etc.[54,55] Silicon, as a centrosymmetric material, has remarkable third-order nonlinearity. THG has been successfully reported in a silicon PCNC, which has potential applications to extend the light-sources in silicon to visible spectral range.[54] The above-demonstrated results indicate the topological edge modes of the SSH structure composed by PCNCs could also support strongly localized electric field, which interestingly has an extra attribute of robustness. It could therefore provide a platform to realize resonance enhanced silicon THG.

To carry out that, a pulsed laser (pulse width=8.8 ps, and repetition rate=18.5 MHz) with tunable wavelength is employed as the pump laser. The scattering THG signal from the structure is collected by the objective lens, which is then separated from the scattering pump laser by a short-pass filter and acquired using a visible spectrometer. By tuning the wavelength of the pump laser to match with the wavelength of the $M_1'$ mode (~1549.05 nm), a strong signal peak is observed at one third of the pump wavelength (~516.35 nm), as shown in **Figure 4**a. With a visible CMOS camera to image the SSH structure through the objective lens, an obvious green light spot is observed at the left terminal nanocavity $A_1$. To verify that the peak originates from the THG process, we examine its intensity dependence on the pump power, as shown in Figure 4b. As the pump power is increased gradually, the THG intensity varies in a cubic function. It shows a fitting line with a slope of ~3.09 in the log-log coordinate system, which is a typical character of the THG process. By fixing the wavelength and power of the pump laser, we also acquire the THG signal from the position shown by the red star of the trivial SSH structure, which is significantly weak, as shown in Figure 4c. With comparison to the resonance enhanced THG peak shown in Fiure 4a, an enhancement factor of 2460 is realized arisen from the topological edge mode.

To confirm that the strongly enhanced THG benefits from the intense localized electric field of the topological edge mode, the dependence of the THG intensities from the SSH structure on the pump wavelength is examined, as plotted in Figure 4d. With a constant pump power, as the pump wavelength is tuned away from the resonance wavelength, THG intensity decreases gradually to a mostly undetectable level. For a cavity resonance mode, the densities of optical power at different wavelengths are governed by a Lorentzian function ($f_{Lorentz}$), as indicated in Figure 2g. In the THG process, THG intensity typically varies as the cubic function of the pump power. Hence, the obtained THG intensities with respect to the pump wavelength should be established by the cubic function of $f_{Lorentz}$, as indicated by the fitting curve $(f_{Lorentz})^3$ in Figure 4d. This clearly indicates that the achievement of strong THG relies on the edge mode.

The edge mode-enabled THG process is further illustrated by implementing its spatial mapping when the SSH structure is spatially moved in the *x-y* plane. Pumped with the on-resonance laser, the scattering THG signal and pump laser are simultaneously monitored using visible and near-infrared photodetectors, respectively. Figure 4e and 4f display the measured results. Only when the focusing spot of the pump laser overlaps with the location of the left terminal nanocavity $A_1$, the edge mode can be excited successfully. This matches well with the simulation result shown in Figure 2i. Because the THG process is a cubic function of the pump light, and it is pumped by the near-field of the resonance edge mode, the spatial area of the THG is much smaller than that of the edge mode. To clarify this, in Figure 4g, we plot the spatial profiles of the THG (red dots) and edge mode (black dots) along the white dashed lines in Figure 4e and 4f. The enhanced THG mainly focuses on the left terminal nanocavity $A_1$ within the strongest field distribution, and its half-height width of the spatial distribution along *x*-axis is reduced by 0.71 μm. In addition, the exponentially decaying tail of edge mode along the chain is restrained by about 10 dB in the THG.

## 4. Conclusion

In conclusion, we have experimentally observed a strongly enhanced THG process in a silicon SSH structure consisting of coupled PCNCs, which is enabled by the localized electric field of the topological edge mode. The edge mode of the SSH structure not only naturally inherits resonant properties of a single PCNC, but also manifests topological feature with steady robustness extending well beyond an individual cavity. From the fabricated SSH structure, the enhancement factor of THG is more than three orders of magnitude with respect to the THG of the trivial SSH structure. Compared with the spatial distribution of the edge mode, the THG exhibits suppression of the decaying tail of the edge mode due to the cubic relationship between its intensity and that of the edge mode. Our results indicate that the SSH structure of PCNCs could provide a promising platform for topology-driven nonlinear processes in device development such as robust visible emitters and entangled photon pairs on silicon chips.

**Acknowledgements**

The Financial support for this work was provided by the Key Research and Development Program (Grant No. 2017YFA0303800), the National Natural Science Foundation (Grant Nos. 91950119, 11634010, 61775183, 61905196), the Key Research and Development Program in Shaanxi Province of China (2020JZ-10), the Fundamental Research Funds for the Central Universities (310201911cx032, 3102019JC008), and the Doctorate Foundation of Northwestern Polytechnical University (CX201924).

Received: ((will be filled in by the editorial staff))
Revised: ((will be filled in by the editorial staff))
Published online: ((will be filled in by the editorial staff))